\documentclass{article}
\usepackage{spconf,amsmath,graphicx}

\usepackage{cite}
\usepackage{amsmath,amssymb,amsfonts}
\usepackage{algorithmic}
\usepackage{graphicx}
\usepackage{textcomp}

\usepackage{amsmath,graphicx}
\usepackage{psfrag}
\usepackage{siunitx}
\usepackage{bm,bbm}

\usepackage{tikz,pgfplots}
\usepackage{tikzscale}

\usepackage{rotating}
\usepackage{wrapfig}
\usepackage[linesnumbered,lined,commentsnumbered, ruled]{algorithm2e}

\usepackage[acronym]{glossaries}

\newacronym{DNN}{DNN}{deep neural network}
\newacronym{CDR}{CDR}{coherent-to-diffuse power ratio}
\newacronym{DoA}{DoA}{direction-of-arrival}
\newacronym{TDoA}{TDoA}{time difference of arrival}
\newacronym{MSE}{MSE}{mean-squared error}
\newacronym{MLP}{MLP}{multilayer perceptron}
\newacronym{CRNN}{CRNN}{convolutional recurrent neural network}
\newacronym[plural=GPs,firstplural=Gaussian processes (GPs)]{GP}{GP}{Gaussian process}
\newacronym{GRU}{GRU}{gated recurrent unit}
\newacronym{RIR}{RIR}{room impulse response}
\newacronym{AWGN}{AWGN}{additive white Gaussian noise}
\newacronym{SNR}{SNR}{signal-to-noise ratio}
\newacronym{STFT}{STFT}{short-time Fourier transform}
\newacronym[plural=PSDs,firstplural=power spectral densities~(PSDs)]{PSD}{PSD}{power sprectral density}
\newacronym{AE}{AE}{absolute error}
\newacronym{MAE}{MAE}{mean-absolute error}
\newacronym{PE}{PE}{position error}
\newacronym{MPE}{MPE}{mean position error}
\newacronym{WASN}{WASN}{wireless acoustic sensor network}
\newacronym{OoR}{OoR}{out-of-range}
\newacronym{ASR}{ASR}{automatic speech recognition}
\newacronym{TDNN}{TDNN}{time delay neural network}
\newacronym{CNN}{CNN}{convolutional neural network}
\newacronym{WLS}{WLS}{weighted least squares}
\newacronym{LS}{LS}{least squares}
\newacronym{RANSAC}{RANSAC}{random sample consensus}

\newacronym{GARDE}{GARDE}{\textbf{G}eometry c\textbf{A}libration f\textbf{R}om \textbf{D}istance \textbf{E}stimates}
\newacronym{MDS}{MDS}{Multi Dimensional Scaling}

\newacronym{CWLS}{CWLS}{constrained weighted least squares}
\newacronym{CRLB}{CRLB}{Cramer-Rao lower bound}
\newacronym{RMSE}{RMSE}{Root Mean Square Error}

\newacronym{CDF}{CDF}{Cumulative distribution function}
\newacronym{CDF2}{CDF}{Cumulative distribution function}

\usepackage[colorlinks]{hyperref}
\usepackage{xcolor}
\usepackage{import}
\usepackage{booktabs}
\usepackage{mathtools}
\usepackage{microtype}

\newcommand{\dE}{\hat{d}}

\newcommand{\dP}{\tilde{d}}

\newcommand{\PES}{\hat{P}}
\newcommand{\PGT}{P}
\newcommand{\PVAR}{\tilde{P}}

\newcommand{\OES}{\hat{O}}
\newcommand{\OGT}{O}
\newcommand{\OVAR}{\tilde{O}}

\newcommand{\norm}[1]{\left\lVert#1\right\rVert_2}

\usepackage[skip=8pt]{caption}
\title{Iterative Geometry Calibration from Distance Estimates \\ for Wireless Acoustic Sensor Networks}
\name{Tobias Gburrek, Joerg Schmalenstroeer, Reinhold Haeb-Umbach}
\address{Department of Communications Engineering, Paderborn University, Germany \\ \{gburrek, schmalen, haeb\}@nt.uni-paderborn.de}

\begin{document}
\ninept
\maketitle
\begin{abstract}
	In this paper we present an approach to geometry calibration in wireless acoustic sensor networks, whose nodes are assumed to be equipped with a compact microphone array.
	The proposed approach solely works with estimates of the distances between acoustic sources and the nodes that record these sources.
	It consists of an iterative weighted least squares localization procedure, which is initialized by multidimensional scaling.
	Alongside the sensor node locations, also the positions of the acoustic sources are estimated.
	Furthermore, we derive the \gls{CRLB} for source and sensor position estimation, and show by simulation that the estimator is efficient.
\end{abstract}
%

\section{Introduction}
In our scenario, the \gls{WASN} hardware consists of  sensor nodes, each equipped with a microphone array. The nodes are at random positions in a room, and they are connected via a WiFi network.
The task to be solved, called geometry calibration, is to estimate the positions of all nodes from the observed acoustic signals in the room. Knowledge of the sensor network's geometry is required, if the \gls{WASN} is to be used to localize speakers, or to enable proximity-based recording and handover functionality \cite{Schmalen2009}.

In general, geometry calibration can be solved by optimizing a cost function that describes the difference between an assumed geometry and one computed from measurements. An overview about different geometry calibration methods can be found in \cite{Plinge16b}.

In previous publications, \gls{DoA} information was employed to solve the geometry calibration task \cite{Schmalenstroeer2011, JaScHa12}. From this, however, no distance information can be obtained, leaving the scaling of the found geometry undetermined. Thus, additional information, e.g., gleaned from another modality, such as video \cite{JaHa2015}, is required  to fix the scaling.

In recent publications \cite{brendel_distributed_2019, brendel_probabilistic_2019, Gburrek2020} it was shown how an estimate of the distance between the acoustic source and the microphone array can be obtained from the recorded audio signal. Distances as inputs allow to employ techniques for geometry calibration which were previously used to localize mobile devices based on signal strength information. In \cite{Cheung04}, for example, a system of equations based on the distance estimates is set up, where the unknown parameter vector consists of the Cartesian coordinates of the mobile device and a range variable which depended on the Cartesian coordinates. Hence, the system of equations is solved approximately by a \gls{CWLS} approach. The authors of \cite{Spirito01} established also a system of equations but subtracted one equation from the system such resulting in an optimization problem that can be solved by unconstrained \gls{LS} without resorting to approximations.
However, both approaches require the knowledge of the base station positions. This would correspond to the knowledge about the acoustic sources' position, which is not available in our scenario.

Besides the \gls{LS} based approaches, also \gls{MDS} based methods have been proposed for location estimation and geometry calibration. The authors of \cite{Cheung05} modified \gls{MDS} for mobile location estimation using time-of-arrival measurements of a signal emitted from the mobile station. In \cite{Amar2010} \gls{MDS} is extended such that it can handle missing distance information by applying a projection. 

In this paper we present a geometry calibration algorithm that uses source-node distance estimates from a \gls{DNN} based acoustic distance estimator \cite{Gburrek2020}. The distance estimation method does not required the nodes of the sensor network to be time synchronized, which is a major advantage over many other geometry calibration approaches.
Our approach combines ideas from mobile phone localization, e.g., \cite{Spirito01}, and iterative optimization techniques. Additionally, we show by simulation that the used position estimator is efficient, since it reaches the \gls{CRLB}.

The paper is organized as follows:
In Sec.~\ref{SEC:GEO} distance-based geometry calibration is explained, followed by the \gls{CRLB} derivation of the variance of the corresponding position estimates in Sec.~\ref{SEC:CRLB}. Implementation issues are discussed in Sec.~\ref{SEC:GARDE}, before experimental results are presented in Sec.~\ref{SEC:EXP} and conclusions are drawn in Sec.~\ref{SEC:CON}.


\clubpenalty = 10000
\widowpenalty = 10000
\displaywidowpenalty = 10000

\section{Geometry calibration}  \label{SEC:GEO}
We assume a setup with $N>3$ sensor nodes at positions $\Omega_{\bm{P}} \coloneqq \{\bm{\PGT}_1, \ldots, \bm{\PGT}_N\}$ and $K>3$ spatially distributed acoustic sources at positions $\Omega_{\bm{O}} \coloneqq \{\bm{\OGT}_1, \ldots, \bm{\OGT}_K  \}$. 
Our geometry calibration procedure, named \gls{GARDE}, assumes that all nodes can observe the same local audio sources and that the node positions remain fixed.

The \gls{GARDE} task can be summarized as follows: 
Based on the estimates $\dE_{n,k}$, i.e., the distances between acoustic sources at unknown positions $\bm{\OGT}_k$ and 
sensor nodes at unknown positions $\bm{\PGT}_n$, the sensor nodes' positions should be inferred. The $\dE_{n,k}$ stem from the \gls{DNN}-based distance estimator that we proposed in~\cite{Gburrek2020}.
\gls{GARDE} delivers estimates of all unknown positions $\Omega=\Omega_{\bm{\PGT}} \cup \Omega_{\bm{\OGT}}$ by solving the following optimization problem:
\begin{align}
\hat{\Omega} = \underset{\Omega}{\operatorname{argmin}}\underbrace{\sum_{k=1}^K \sum_{n=1}^N \left(\dE_{n,k}^{\: 2} - \norm{\bm{\PGT}_{n}-\bm{\OGT}_{k}}^2 \right)^2}_{J(\Omega)}.
\label{eq:costfunction}
\end{align}
\vfill

Calculating the gradient of $J(\Omega)$ w.r.t.\ the $g$-th unknown position $\bm{\PGT}_{g}$ and setting the result equal to zero gives:
\begin{eqnarray}
\small
\frac{\partial  J}{\partial \bm{\PGT}_{g}}\Bigr|_{\hat{\Omega}} \stackrel{!}{=} 0 \! \Leftrightarrow \!
\sum_{k=1}^K \left(\bm{\PES}_{g}\!-\!\bm{\OES}_{k}\right) (\underbrace{ \dE_{g,k}^{\:2} \!-\! \norm{\bm{\PES}_{g}\!-\!\bm{\OES}_{k}}^2}_{e_{gk}}) = 0.\!
\label{eq:deriverative}
\end{eqnarray}
Similarly, for the gradient of $J(\Omega)$ w.r.t.\ the $h$-th unknown acoustic source position $\bm{\OGT}_h$ it follows:
\begin{eqnarray}
\footnotesize
\frac{\partial J}{\partial \bm{\OVAR}_h}\Bigr|_{\hat{\Omega}}  \stackrel{!}{=} 0  \!\Leftrightarrow  \!
\sum_{n=1}^N \left(\bm{\PES}_{n} \!- \!\bm{\OES}_{h}\right)(\underbrace{ \dE_{n,h}^{\: 2} \! -  \!\norm{\bm{\PES}_{n} \!- \!\bm{\OES}_{h}}^2}_{e_{nh}}) = 0. \!
\label{eq:deriverative2}
\end{eqnarray}

For \eqref{eq:deriverative} and \eqref{eq:deriverative2} no closed form solution exists. However, minimizing alternatingly the $e_{gk}$ and $e_{nh}$ should at least minimize the cost function. To this end, an iterative  \gls{WLS} approach can be set up, assuming either the positions $\bm{\OES}_k$ in \eqref{eq:deriverative} or the positions $\bm{\PES}_n$ in \eqref{eq:deriverative2} to be fixed.
The similarity of \eqref{eq:deriverative} and \eqref{eq:deriverative2} allows to state a common solution for both problems.

Following the ideas of \cite{Spirito01}, the localization procedure has to select one sensor node as reference node, which has to be chosen judiciously, as the resulting localization error depends on the precision of the distance estimate of the selected sensor. Here, we always select the node with the smallest distance estimate to the acoustic source as reference node, as it was shown in \cite{Gburrek2020} that the distance estimation error increases with the distance between audio source and sensor. Informal experiments showed an overall error reduction by approximately a factor of two compared to randomly selecting the reference node.

Let us assume that for some $\bm{\PGT}_{\nu}$ $\dE_{\nu,i}{\leq}\dE_{j,i} \:\: \forall i,j$ holds and
that $\bm{\PGT}_{\nu}$ is chosen to be the center of the used coordinate system ($\bm{\PVAR}_{n}=\bm{\PGT}_{n}-\bm{\PGT}_{\nu}$, $\bm{\OVAR}_{k}=\bm{\OGT}_{k}-\bm{\PGT}_{\nu}$). For this choice we get:
\begin{align} \label{EQ:radius}
\left(\PVAR_{n,x} - \OVAR_{k,x}\right)^2 + \left(\PVAR_{n,y} - \OVAR_{k,y}\right)^2 = \dE_{n,k}^{\: 2}  &  \:\: \:\:\:\: \:\: \forall n \neq \nu \\ \label{EQ:Refradius}
\OVAR_{k,x}^2 + \OVAR_{k,y}^2 = \dE_{\nu,k}^{\: 2}  &  \:\: \:\:\:\: \:\: n = \nu  
\end{align}
with $n \in \{1,\ldots, N\}$.
Subtracting \eqref{EQ:Refradius} from all equations of \eqref{EQ:radius} and stacking the resulting equations in a 
matrix formulation (excluding the $\nu$-th one) gives:
\begin{align} \nonumber
\underbrace{\left[\begin{matrix}
	2\PVAR_{1,x} & 2\PVAR_{1,y} \\ \vdots & \vdots \\ 2\PVAR_{n,x} & 2\PVAR_{n,y} \\
	\end{matrix} \right]}_{\bm{R}}
\left[\begin{matrix}
\OVAR_{k,x} \\ \OVAR_{k,y}
\end{matrix} \right] = 
\underbrace{\left[\begin{matrix}
	\dE_{\nu,k}^{\: 2} + \PVAR_{1,x}^2 + \PVAR_{1,y}^2 - \dE_{1,k}^{\: 2} \\ \vdots \\
	\dE_{\nu,k}^{\: 2} + \PVAR_{n,x}^2 + \PVAR_{n,y}^2 - \dE_{N,k}^{\: 2}
	\end{matrix} \right]}_{\bm{b}}
\end{align}
The \gls{DNN}-based distance estimates are corrupted by an error which tends to be heteroscedastic (see~\cite{Gburrek2020}). So \gls{WLS} is employed to estimate the location $\bm{\OES}_k$ with
\begin{align} 
\bm{\OES}_k = \left( \bm{R}^T \bm{W}\bm{R}\right)^{-1} \bm{R}^T \bm{W} \bm{b} + \bm{\PGT}_{\nu}, \label{EQ:WLS}
\end{align}
where the weighting matrix elements are chosen to $W_{n,n} = 1/\dP_{n,k}^{\: 2}$, reflecting the earlier mentioned observation that small distances exhibit a lower estimation error.
The concrete definition of the distances $\dP_{n,k}$ can be found in Sec. \ref{SEC:GARDE}.

The \gls{WLS} solution from \eqref{EQ:WLS} can also be used to estimate the position of the sensor nodes from the distance estimates by exchanging the roles of $\bm{\PES}$ and $\bm{\OES}$.
This results in an iterative positioning procedure used in \gls{GARDE}, as it will be explained in Sec.~\ref{SEC:GARDE}. 
However, an initial guess is needed to start the iterative algorithm which can be either positions of the nodes or the observations. We propose to use a \gls{MDS}-based initialization for the sensor node positions, which we present in the following.

\subsection{MDS-based Initialization}
\begin{figure}[htb]
	\hspace{1.5cm}
	\def\svgwidth{0.85\columnwidth}
	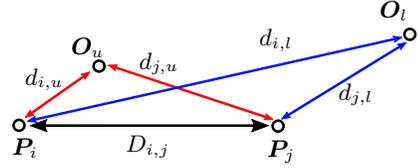
	\caption{Inter-node distance estimation example using subtraction (blue distances) and addition (red distances) of node to acoustic source distances for lower and upper bound approximation (see \eqref{eq:triang}).}
	\label{fig:upperlower}
\end{figure}

\gls{MDS} employs a distance matrix $\bm{D}$ containing all inter-node distances $D_{i,j}$ with $i,j\in[1,N]$ to estimate the positions of the senor nodes.
But, the utilized \gls{DNN}-based distance estimator delivers only a set of distances $\bm{A}_{P} = \{\dE_{1,1},\ldots,\dE_{N,K}\}$ between unknown positions of nodes $\bm{\PGT}_n$ and unknown positions of sources $\bm{\OGT}_k$. Neither distance estimates between sensors nor distance estimates between sources are available. To overcome this difficulty, we utilize the triangular inequality (compare Fig.~\ref{fig:upperlower})
\begin{align}
\max_l(|\dE_{i,l}-\dE_{j,l}|) \leq  D_{i,j} \leq \min_u(\dE_{i,u}+\dE_{j,u}),
\label{eq:triang}
\end{align}
where $i,j$ are sensor node indices and $u,l$ are acoustic source  indices. 
The bounds can be used to obtain a first estimate for the distance between the nodes $i$ and $j$:
\begin{align}
\widehat{D}_{i,j} = [\max_l(|\dE_{i,l}-\dE_{j,l}|) + \min_u(\dE_{i,u}+\dE_{j,u})]/2.
\end{align}
Based on $\widehat{\bm{D}}$ an initial geometry is estimated via classical \gls{MDS} as described in \cite{BorgGroenen2005}.


\section{\acrlong{CRLB}} \label{SEC:CRLB}
\begin{figure}[htb]
	\centering
	\includegraphics[width=\columnwidth]{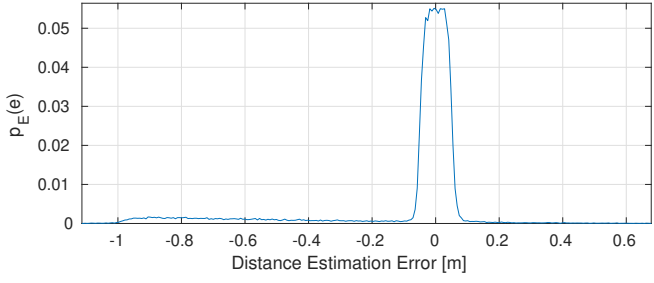}
	\caption{Histogram of distance estimation errors from $80.000$ observations.}
	\label{Fig:errorhist}
\end{figure}

In this section a \gls{CRLB} for the \gls{GARDE} approach is derived, following the ideas of \cite{McGuire00} and \cite{SchHaeb16}. The \gls{CRLB} for the \gls{GARDE} position estimator, including acoustic source and sensor node positions, can be stated as follows. The observation errors $e_{n,k}$ which are given by
\begin{align}
e_{n,k} = \dE_{n,k} - \norm{\bm{\PGT}_{n}-\bm{\OGT}_{k}}
\end{align}
are assumed to follow a zero-mean Gaussian with variance $\sigma_{d}^2$, compare Fig.~\ref{Fig:errorhist}.

Given the position $\bm{\OGT}_{k}$, the  distribution of the $N$ observations  $\bm{\dE}_k = \left[\begin{matrix} \dE_{1,k} &\cdots &\dE_{N,k} \end{matrix} \right]^T$ is
\begin{align}
p\left(\bm{\dE}_k|\bm{\OGT}_{k}\right) = \xi \cdot \exp\left\{ \sum\limits_{n=1}^{N} \frac{-\left(\dE_{n,k} - \dP_{n,k} \right)^2}{2 \sigma_{d}^2}  \right\},
\end{align}
with constant $\xi = ((2\pi)^{\frac{N}{2}} \prod\limits_{n=1}^{N} \sigma_{d})^{-1}$ and  the ground truth distance from positions
\begin{small}
\begin{eqnarray*}
\dP_{n,k} = \norm{\bm{\PGT}_{n}-\bm{\OGT}_{k}} = \sqrt{(\PGT_{n,x}-\OGT_{k,x})^2 + (\PGT_{n,y}-\OGT_{k,y})^2}.
\end{eqnarray*}
\end{small}
The \gls{CRLB} for the position estimate of $\bm{\OGT}_{k} = \left[\begin{matrix}\OGT_{k,x} &\OGT_{k,y}\end{matrix} \right]^T$ is
derived from the log-likelihood $L_k$ with
\begin{eqnarray}
L_k := \mathtt{ln} \left( p(\bm{\dE}_k|\bm{\OGT}_{k})\right) = \mathtt{ln}(\xi) - \sum\limits_{n=1}^{N} \frac{\left(\dE_{n,k} - \dP_{n,k}\right)^2}{2 \sigma_{d}^2}.  
\end{eqnarray}
The first and second order derivatives w.r.t.\ $\OGT_{k,x}$ of the log-likelihood are:
\begin{small}
\begin{align}
&\frac{\partial \: L_k}{\partial \OGT_{k,x}}   = 
\sum\limits_{n=1}^{N} \frac{1}{\sigma_{d}^2} \left(\dE_{n,k} - \dP_{n,k}\right) 
\frac{\partial}{\partial \OGT_{k,x}} \left(\dP_{n,k}\right)  \\\label{EQ:SecOrderDer}
&\frac{\partial^2 \: L_k }{(\partial \OGT_{k,x})^2}  = 
\sum\limits_{n=1}^{N} \frac{-1}{\sigma_{d}^2} 
\left[ 
\left(\frac{\partial \: \dP_{n,k}}{\partial \OGT_{k,x}} \right)^2 -
\left(\dE_{n,k} - \dP_{n,k}\right) \frac{\partial^2 \left(\dP_{n,k}\right)}{(\partial \OGT_{k,x})^2} 
\right] 
\end{align}
\end{small}
Applying the expectation operator to \eqref{EQ:SecOrderDer} results in
\begin{align}
E\left\{ \frac{\partial^2  \: L_k}{(\partial \OGT_{k,x})^2} \right\} = \sum\limits_{n=1}^{N} \frac{-1}{\sigma_{d}^2} 
\frac{(\PGT_{n,x}-\OGT_{k,x})^2}{\norm{\bm{\PGT}_{n}-\bm{\OGT}_{k}}^2}
\end{align}
Similarly, the derivatives w.r.t.\ $\OGT_{k,y}$ can be found. For a short hand notation we summarize the terms by:
\begin{small}
\begin{align}
E\left\{ \frac{\partial^2  \: L_k}{(\partial \OGT_{k,x})^2} \right\} = \sum\limits_{n=1}^{N} \frac{-1}{\sigma_{d}^2} 
\frac{(\PGT_{n,x}-\OGT_{k,x})^2}{\norm{\bm{\PGT}_{n}-\bm{\OGT}_{k}}^2} \coloneqq \gamma_{xx}\label{EQ:GAMMAXX} \\ 
E\left\{ \frac{\partial^2  \: L_k}{(\partial \OGT_{k,y})^2} \right\} = \sum\limits_{n=1}^{N} \frac{-1}{\sigma_{d}^2} 
\frac{(\PGT_{n,y}-\OGT_{k,y})^2}{\norm{\bm{\PGT}_{n}-\bm{\OGT}_{k}}^2} \coloneqq \gamma_{yy}\label{EQ:GAMMAYY} \\\nonumber 
E\left\{ \frac{\partial^2  \: L_k}{\partial \OGT_{k,x} \: \partial \OGT_{k,y}} \right\} = E\left\{ \frac{\partial^2  \: L_k}{\partial \OGT_{k,y} \: \partial \OGT_{k,x}} \right\} \\
= - \sum\limits_{n=1}^{N} \frac{1}{\sigma_{d}^2} 
\frac{(\PGT_{n,x}-\OGT_{k,x})(\PGT_{n,y}-\OGT_{k,y})}{\norm{\bm{\PGT}_{n}-\bm{\OGT}_{k}}^2} \coloneqq \gamma_{xy} = \gamma_{yx} \label{EQ:GAMMAXY}
\end{align}
\end{small}
Finally, the \gls{CRLB} can be stated by
\begin{align*}
\mathtt{CRLB}(\bm{\OGT}_{k}) = - \left[E \left\{
\begin{matrix}
\frac{\partial^2L_k}{\partial \OGT_{k,x}\partial \OGT_{k,x}} &
\frac{\partial^2 L_k}{\partial \OGT_{k,x}\partial \OGT_{k,y}} \\
\frac{\partial^2L_k}{\partial \OGT_{k,y}\partial \OGT_{k,x}} &
\frac{\partial^2 L_k}{\partial \OGT_{k,y}\partial \OGT_{k,y}} \\
\end{matrix}
\right\} \right]^{-1}, 
\end{align*}
whereby the variance of the estimate $\bm{\OES}_{k}$ is lower bounded by
\begin{small}
\begin{align*}
\mathtt{var}(\bm{\OES}_{k,x}) \geq \frac{\gamma_{yy}}{\gamma_{xy}^2 - \gamma_{xx} \gamma_{yy}}  \:\text{ and }\:
\mathtt{var}(\bm{\OES}_{k,y}) \geq \frac{\gamma_{xx}}{\gamma_{xy}^2 - \gamma_{xx} \gamma_{yy}}.
\end{align*}
\end{small}
The \gls{CRLB} enables the estimation of a lower bound for the \gls{RMSE} value  of position $\bm{\OES}_{k}$ with
\begin{align}
\mathtt{RMSE}(\bm{\OES}_{k}) \geq \sqrt{\frac{\gamma_{xx} + \gamma_{yy}}{\gamma_{xy}^2 - \gamma_{xx} \gamma_{yy} }}.
\label{EQ:RMSE}
\end{align}
Since the distances $\dE_{n,k}$ are used both for estimating the positions of sensors and acoustic sources, the result of \eqref{EQ:RMSE} can be easily adapted to obtain a lower bound for $\mathtt{RMSE}(\bm{\PES}_{n})$ by exchanging the sum index from $n$ to $k$ in $\gamma_{xx}$, $\gamma_{xy}$ and $\gamma_{yy}$ in \eqref{EQ:GAMMAXX}-\eqref{EQ:GAMMAXY}.


\section{GARDE Implementation details}\label{SEC:GARDE}
The \gls{GARDE} algorithm for geometry calibration is summarized in Alg.~\ref{GARDE}. Lines 2-5 correspond to the estimation of initial values of the nodes' positions using \gls{MDS}.
Based on these initial sensor position estimates the \gls{WLS} localization is used to estimate the positions $\bm{\OES}$ of the acoustic sources in line 6.
To this end, the generic function $\widehat{\Omega}_{\bm{a}} = \mathtt{LSPos}(\widehat{\Omega}_{\bm{b}} , \bm{A}_{P}, \dP_{n,k})$ utilizes the distance observations $\bm{A}_{P}$ and sensor node positions $\widehat{\Omega}_{\bm{b}}=\widehat{\Omega}_{\bm{P}}$ to get source position estimates $\widehat{\Omega}_{\bm{a}}=\widehat{\Omega}_{\bm{O}}$.
Thereby, $\widehat{\Omega}_{\bm{a}} = \mathtt{LSPos}(\widehat{\Omega}_{\bm{b}} , \bm{A}_{P}, \dP_{n,k})$ utilizes \eqref{EQ:WLS} with the weightening distances $\dP_{n,k}$ to determine each element of $\widehat{\Omega}_{\bm{a}}$ separately.

These first position estimates are refined by additional \gls{LS} iterations in line 7.
Subsequently, the iterative estimation, including annealing steps, is given by the for-loops.
Annealing the variance parameter $\mu(g)$ turned out to be important to overcome unfavorable local optima.
After finishing an iteration the newly found positions $\widehat{\Omega}$ are compared with the previous best estimate $\widehat{\Omega}_{pos, o}$ via the $\mathtt{OptSelect}$ function in line 11 with $\widehat{\Omega}_{o} = \mathtt{OptSelect}(\widehat{\Omega}, \widehat{\Omega}_{o}, \bm{A}_{P})$.
The function selects the geometry with the smallest average error between the observed distances $\bm{A}_{P}$ and the distances calculated from the estimated node and acoustic source positions. 

\begin{algorithm}[htb]
	\small
	\KwData{Observed distances $\bm{A}_{P}$\;}
	Init: $\alpha=0.2$; $\beta=0.2$\;
	\For{$i = 1 \rightarrow N$, $j = 1 \rightarrow N$ }{
		$\widehat{D}_{ij} = \frac{1}{2} \max\limits_l(|\dE_{il}-\dE_{jl}|) + \frac{1}{2} \min\limits_k(\dE_{iu}+\dE_{ju})$
	}
    $\widehat{\Omega}_{\bm{\PGT}} = \mathtt{MultiDimensionalScaling}(\bm{\widehat{D}})$\;
	$\widehat{\Omega}_{\bm{\OGT}} = \mathtt{LSPos}\left(\widehat{\Omega}_{\bm{\PGT}} , \bm{A}_{P}, \dE_{n,k} \right)$\;
	$\widehat{\Omega} = \mathtt{Iterate}(\widehat{\Omega}, \bm{A}_{P})$\;
	$\widehat{\Omega}_{o} = \widehat{\Omega}$\;
	\For{$g = 1 \rightarrow \mathtt{NumAnnealing}$}{
		$\widehat{\Omega} = \mathtt{Iterate}(\widehat{\Omega}, \bm{A}_{P})$\;
		$\widehat{\Omega}_{o} = \mathtt{OptSelect}(\widehat{\Omega}, \widehat{\Omega}_{o}, \bm{A}_{P})$\;
		$\bm{\PES}_n = \bm{\PES}_{n,o} + \mu(g) \cdot \mathtt{randn}(), \ \forall n \in \{1, \ldots, N\}$\;
		$\bm{\OES}_k = \bm{\OES}_{k,o} + \mu(g) \cdot \mathtt{randn}(), \ \forall k \in \{1, \ldots, K\}$\;}	
	\KwResult{$\widehat{\Omega}_{o}$;}	
	\SetKwFunction{FIter}{Iterate}
	\SetKwProg{Fn}{Function}{:}{}
	\Fn{\FIter{$\widehat{\Omega}$, $\bm{A}_{P}$}}{
		\For{$i = 1 \rightarrow \mathtt{NumIterations}$}{
			$\widehat{\Omega}_{\bm{\OGT}, fit} = \mathtt{FitSelect}(\widehat{\Omega}, \bm{A}_{P})$\;
			$\widehat{\Omega}_{\bm{\PGT}, fit} = \mathtt{LSPos}(\widehat{\Omega}_{\bm{\OGT}, fit} , \bm{A}_{P}, ||\bm{\PES}_{n}-\bm{\OES}_{k}||_2)$\;
			$\widetilde{\Omega}_{\bm{\PGT}, fit} = \mathtt{Map2Ref}(\widehat{\Omega}_{\bm{\PGT}, fit}  \rightarrow\widehat{\Omega}_{\bm{\PGT}})$\;
			$\bm{\PES}_n = \alpha \cdot \bm{\PES}_n + (1-\alpha) \cdot \widetilde{\bm{\PGT}}_{n,fit}, \ \forall n \in \{1, \ldots, N\}$\;
			$\widetilde{\Omega}_{\bm{\OGT}} = \mathtt{LSPos}(\widehat{\Omega}_{\bm{\PGT}}, \bm{A}_{P}, \dE_{n,k})$\;
			$\bm{\OES}_k = \beta \cdot \bm{\OES}_k + (1-\beta) \cdot \widetilde{\bm{\OGT}}_k, \ \forall k \in \{1, \ldots, K\}$\;
		}
		\KwRet $\widehat{\Omega}$;
	}
\caption{\gls{GARDE} algorithm} \label{GARDE}
\end{algorithm}

In each iteration two auxiliary functions are used.
The function $\widetilde{\Omega}_{\bm{P}, fit} = \mathtt{Map2Ref}(\widehat{\Omega}_{\bm{P}, fit}  \rightarrow \widehat{\Omega}_{\bm{P}})$ maps the set of positions $\widehat{\Omega}_{\bm{P}, fit}$ to a set of reference positions $\widehat{\Omega}_{\bm{P}}$ via rotation and translation operations.
This processing step is necessary to enable the merge steps in line 20 and 22, where the results of the current iteration are combined with results of previous iterations.
The second function $\widehat{\Omega}_{\bm{O}, fit} = \mathtt{FitSelect}(\widehat{\Omega}, \bm{A}_{P})$ selects a set of observations $\widehat{\Omega}_{\bm{O}, fit}$ from $\widehat{\Omega}_{\bm{O}}$ that best fit the distances $\bm{A}_{P}$ and the assumed sensor positions $\widehat{\Omega}_{\bm{}}$ in terms of minimum average error.


\section{Experiments} \label{SEC:EXP}
For our experiments we utilize the image source method~\cite{image_source} from the implementation of \cite{habets_rir} to simulate \glspl{RIR} for $40$ different setups with randomly placed nodes ($4$ nodes in a room) and 500 observations.
The reverberation time $T_{60}$ is selected to be $\SI{200}{ms}$ or $\SI{400}{ms}$. The acoustic sources are simulated using speech signals from the TIMIT database \cite{timit}, which are convolved with the \glspl{RIR} to obtain the microphone signals. The \gls{DNN} approach from \cite{Gburrek2020} is used to estimate the distance between sensor node and acoustic source.

\begin{figure}[htb]
	\centering
	\includegraphics[width=0.8\columnwidth]{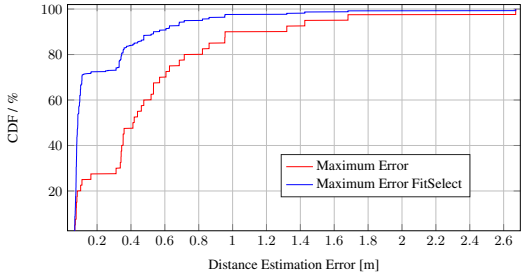}
	\caption{Cumulative distribution function (CDF) of maximum distance error in $\bm{A}_p$ and  selected subset of $\bm{A}_p$ by FitSelect function.}
	\label{Fig:maxerrorsel}
\end{figure}

In Fig.~\ref{Fig:maxerrorsel} the CDF of the maximum distance error in $\bm{A}_p$ is compared against the maximum error after $30$ iterations and annealing rounds for the fitness based selected observations (Alg.~\ref{GARDE}, line 10), i.e., the set of observations used to estimate the final geometry. The experiment shows that \gls{GARDE} is able to identify outliers and to significantly reduce the maximum distance error.

\begin{figure}[htb]
	\centering
	\includegraphics[width=0.9\columnwidth]{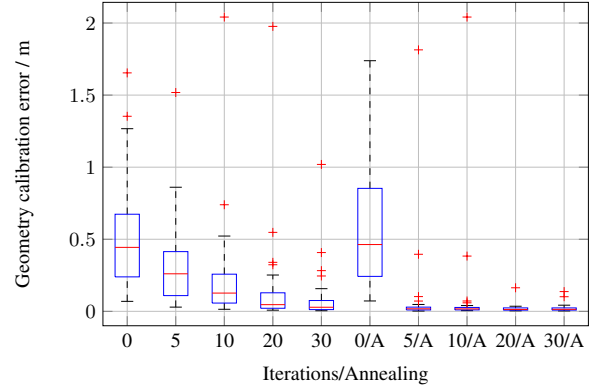}
	\caption{Geometry calibration error for different number of iterations and optionally $30$ annealing rounds (results marked with ''/A'' ).}
	\label{Fig:genit2}
\end{figure}

Simulated annealing is  part of \gls{GARDE} to overcome local minima.  Its effect is illustrated in Fig.~\ref{Fig:genit2}. For different number of iterations ($0 \rightarrow 30$), results with activated annealing (''A'') and without annealing are compared. Overall, annealing significantly reduces the average geometry calibration error, i.e., the minimum \gls{RMSE} between estimated geometry and ground truth after applying an optimal mapping including translation and rotation operations. The depicted results with $0$ iterations reflect the performance of the \gls{MDS} initialization. 

\begin{table}[htb]
\centering
	\caption{\gls{RMSE} of the sensor positions for different geometry calibration methods}
	\begin{tabular}{ c  c  c }
		Method & $T_{60} = \SI{200}{\milli \second}$  & $T_{60} = \SI{400}{\milli \second}$ \\
		\toprule
		\gls{DoA} \cite{JaScHa12} + Scaling \cite{Gburrek2020} & \SI{0.043}{m} & \SI{0.103}{m} \\
		\gls{GARDE} & \SI{0.017}{m} & \SI{0.032}{m} \\
	\end{tabular}
	\label{tab:comparison}
\end{table}
Tab. \ref{tab:comparison} compares \gls{GARDE} ($30$ iterations and annealing rounds) with the \gls{DoA}-based geometry calibration method from \cite{JaScHa12}.
The latter is combined with the scaling approach, we have presented in \cite{Gburrek2020}. \gls{GARDE} outperforms the \gls{DoA}-based method and is less affected by reverberation.

\begin{figure}[htb]
	\centering
	\includegraphics[width=0.9\columnwidth]{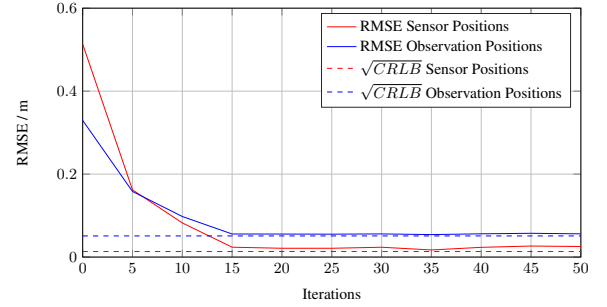}
	\caption{Comparision between \gls{CRLB} of estimator and \gls{RMSE} of observation and sensor positions. Number of iterations and annealing rounds were equally chosen.}
	\label{Fig:genit3}
\end{figure}

In a final experiment the \glspl{RMSE} of the sensor and acoustic source position estimates are compared against the  \gls{RMSE} bounds predicted by the \gls{CRLB}. Our estimator reaches for both the predicted bounds as shown in Fig.~\ref{Fig:genit3}.

\vspace{-0.5cm}
\section{Conclusion}  \label{SEC:CON}
In this paper we have presented the \gls{GARDE} algorithm, an iterative \gls{WLS}-based algorithm for geometry calibration of \glspl{WASN}.
\gls{GARDE} estimates the sensor nodes' positions and the positions of acoustic sources based on distance estimates, which are derived from the recorded audio signals.
Furthermore, \glspl{CRLB} for the geometry calibration approach are derived and compared against the simulation results showing the promising precision of the \gls{GARDE}\footnote{The Python software code of GARDE is available in the paderwasn repository, \url{https://github.com/fgnt/paderwasn}.} algorithm.

\paragraph*{Acknowledgment}
Funded by the Deutsche Forschungsgemeinschaft (DFG, German Research Foundation) - Project 282835863.

\bibliographystyle{IEEEbib}
\bibliography{refs,library,asn_publications,p1,eulib}

\end{document}